\documentclass[smallextended]{svjour3}       
\smartqed  
\usepackage{graphicx}
\usepackage{amssymb} 
\usepackage{amsmath}    
\usepackage{latexsym}   
\usepackage{bm}
\usepackage{times}

\newcommand{\half}{\textstyle{\frac{1}{2}}}

\newcommand{\abr}{\mathrel{\stackrel{\longrightarrow}{A\!B}}\!}

\journalname{Quantum Information Processing}
\begin{document}

\title{Unpredictability and the transmission of numbers}


\titlerunning{Unpredictability and the transmission of numbers}        

\author{John M. Myers         \and
        F. Hadi Madjid 
}


\institute{John M. Myers \at
              Harvard School of Engineering and Applied Sciences, Cambridge, MA
              02138 USA\\
              Tel.: 617-495-5263\\
              \email{myers@seas.harvard.edu}           
           \and
           F. Hadi Madjid \at
           82 Powers Road, Concord, MA 01742 USA\\
           Tel.: 978-349-1808\\
           \email{gmadjid@aol.com}
}

\date{Received: date / Accepted: date}

\maketitle

\begin{abstract}
Curiously overlooked in physics is its dependence on the transmission of
numbers. For example the transmission of numerical clock readings is implicit
in the concept of a coordinate system.  The transmission of numbers and other
logical distinctions is often achieved over a computer-mediated communications
network in the face of an unpredictable environment.  By unpredictable we mean
something stronger than the spread of probabilities over given possible outcomes,
namely an opening to unforeseeable possibilities.  Unpredictability, until now
overlooked in theoretical physics, makes the transmission of numbers
interesting.  Based on recent proofs within quantum theory that provide a
theoretical foundation to unpredictability, here we show how regularities in
physics rest on a background of channels over which numbers are transmitted.

As is known to engineers of digital communications, numerical transmissions
depend on coordination reminiscent of the cycle of throwing and catching by
players tossing a ball back and forth.  In digital communications, the players
are computers, and the required coordination involves unpredictably adjusting
"live clocks" that step these computers through phases of a cycle.  We show how
this phasing, which we call \emph{logical synchronization}, constrains
number-carrying networks, and, if a spacetime manifold in invoked, put
``stripes'' on spacetime.  Via its logically synchronized channels, a network
of live clocks serves as a reference against which to locate events.  Such a
network in any case underpins a coordinate frame, and in some cases the direct
use of a network can be tailored to investigate an unpredictable environment.
Examples include explorations of gravitational variations near Earth.

\end{abstract}

\section{Introduction} \label{sec:1}
Some of us, especially on the theory side, entered physics to evade surprises,
to find the enduring.  What, though, if the enduring is the prevalence of
unpredictable surprises?  By unpredictable we mean something stronger than
uncertain.  While \emph{uncertainty} pertains to a spread in a probability
measure over a given set of possibilities, \emph{unpredictability} allows for
the emergence of \emph{unforeseen possibilities}.  Unpredictability as
discussed here stems from a distinction between evidence and explanations of
that evidence.  This distinction is reflected within quantum theory.  On the
blackboard of theory, given evidence (from the workbench, so to speak) is
expressed as a probability measure over a set of outcomes.  The probability
measure is parametrized by ``knob settings,'' thought of as under experimental
control \cite{tyler07}.  Explanations or predictions of the probabilities are
expressed on the blackboard in terms of wave functions or density operators
representing a prepared quantum state, along with linear operators expressing a
measurement procedure.  We proved that for any given parametrized probability
measure there are infinitely many quantum explanations that generate the given
probabilities, but that disagree with each other about probabilities from
evidence of future experiments, not yet on hand \cite{aop14}.  Choosing a
quantum explanation of given evidence therefore requires a reach beyond logic
to make a guess.  Because of the need to guess, the explanation is not merely
uncertain, but is unpredictable: the guess announces an unforeseen possibility.

Here we offer networks of numerical transmission as structures recognizing the
role of unpredictability in the concept of location.  To locate something one
needs a background against which to locate it.  Usually in physics that
background is a coordinate frame, which for theoretical purposes is represented
by a coordinate system. (One distinguishes a coordinate system as a
mathematical construction from its realization as a coordinate frame involving
measurement uncertainties \cite{soffel03}.)  In Sec.\ \ref{sec:2} we review how
coordinate frames depend on transmissions of numerical clock readings through
an unpredictable environment, so that a location is specified by its relation a
background consisting of numerical clock readings and number-carrying signals,
for example in the Global Positioning System (GPS).  In the theories of special
and general relativity, a coordinate system entails numerically expressed
reference patterns clock readings, toward with one tries to steer a physical
frame.  In general relativity, the possible reference patterns of clock
readings and number-bearing signals are constrained by a metric tensor field.
With the high precision involved in the search for gravitational radiation,
this metric tensor field can be unpredictable, thus subjecting any reference
pattern of clock readings to unpredictability \cite{aop14}.  This
unpredictability is essential to the concepts of \emph{live clocks} and their
\emph{logical synchronization} reviewed in Sec.\ \ref{sec:3}.  The most precise
frames require facing the unpredictability of the reference pattern by
replacing the traditional use of a prescribed reference pattern with one that
is provisional and continually adapted.  Thus at a level of feedback above the
steering toward a given reference pattern, the live clocks of a network detect
and respond to unpredictable failures to steer within an allowable tolerance of
whatever reference pattern is invoked at the moment, followed by employing a
revised reference pattern.

To show how logical synchronization differs from the synchronization
defined by Einstein, in Sec.\ \ref{sec:4} we invoke the assumption of a
spacetime manifold, indeed a flat spacetime,  to show logical
synchronization in a case of live clocks fixed to a rotating platform, where
Einstein synchronization is precluded.  It is noted, however,
that live-clock networks as a concept make no assumption of a spacetime manifold;
spacetime coordinates enter as an optional ingredient in the planning of some,
but by no means all, reference patterns.

Sec.\ \ref{sec:5} summarizes the overall perspective and points to some open
questions concerning the locating of events not by coordinates but in terms of
number-carrying communications channels arising or constructed in particular
situations and dependent on active maintenance.  Such networks locate events in
terms of "who's in touch with whom" over channels linking live clocks that
compute their own rate adjustments in response to measured deviations from a
reference pattern of channels, that itself varies unpredictably.  Indeed, the
needs for changing the reference pattern give evidence for an unpredictable
environment as a topic of experimental investigation.

\section{Transmitted numbers in the theory of reference frames}\label{sec:2}
Quantum theory presupposes coordinate systems as mathematical constructions
that one relates to physical systems.  Coordinate systems depend (at least
locally) on Einstein's imagined patterns of light signals propagating between
imagined \emph{proper} clocks.  In terms of these clocks and signals Einstein
defined the synchronization of proper clocks fixed to a non-rotating, rigid
body in free fall (i.e., a Lorentz frame) and co-defined ``time'' as the
readings of such proper clocks, with the implications that distance from proper
clock $A$ to proper clock $B$ is defined, as in radar, in terms of the duration
at $A$ from the transmission of a light signal to the return of its echo from
$B$.  Specifically, according to Einstein's definition of the synchronization
of proper clocks \cite{einstein05}, clock $B$ is \emph{synchronous} to clock
$A$ if at any $A$-reading $t_A$, $A$ could send a signal reaching $B$ at
$B$-reading $t_B$, such that an echo from $B$ would reach $A$ at $A$-reading
$t'_A$, satisfying the criterion
\begin{equation}\label{eq:es}
t_B=\half(t_A+t'_A).  
\end{equation}

By postulate, proper clocks are free of drift in frequency, so that the
relation that defines synchronization can be thought of as what ``would hold''
if signals were transmitted, without requiring actual transmission.  But when
we turn from coordinate systems as mathematical entities to their realization
by physical coordinate frames, drift of physical clocks, stemming from quantum
uncertainty and other causes, has to be dealt with \cite{aop14}.  Dealing with
it entails attending to the actual transmission of numerical clock readings.
Inspired by computer-mediated digital communications systems, we reflect the
need for the transmission of physical clock readings into theory by representing
a real-time process-control computer that takes part in a network as a modified
Turing machine stepped by a clock \cite{turing}.  To deal with communications
between Turing machines, it is necessary for the clock that steps a Turing
machine to tick at a rate that can be adjusted by commands issuing from that
machine.  We call such an adjustable clock in combination with the Turing
machine that regulates its rate a \emph{live clock} \cite{04345}.

Seen from the standpoint of live clocks as actors in a network, the story of
signaling implicit in Einstein's definition of synchronization (\ref{eq:es})
can be retold as follows.  A live clock $A$ transmits a signal conveying the
very $A$-reading $t_A$ at which the transmission occurs.  Live lock $B$
receives the number-bearing signal at $B$-reading $t_B$ and echoes back a
signal, which conveys the number $t_B$, to live clock $A$, whereupon live
clock $A$ records its $A$-reading $t'_A$ at its receipt of the number $t_B$
from $B$.  Notice that numerical clock readings are transmitted from live clock
to live clock, and that a live clock takes numerical readings of itself at the
transmission and at the receipt of signals that convey clock readings.  The
reading of say clock $B$ at the receipt of a signal carrying a clock reading of
$A$ is distinct from the $A$-reading received.

A network of live clocks acts in response to deviations of relations among its
clock readings from an imagined reference.  One might suppose that the live
clocks of a network could employ a reference stated purely in terms of desired
relations among the numerical clock readings at their transmissions and
receptions of signals. This supposition, however, is wrong, because those
relations provide no scale.  A reference consisting of the Einstein
synchronization relation illustrates the issue of scale.  According to general
relativity, these (blackboard) relations are progressively more precisely
realizable as the clocks over a region increase their tick rates and
correspondingly shrink their separations.  Thus the synchronization relations
have to be augmented by some local scale.  In the International System of Units
(SI), this scale is chosen to be a resonance of cesium 133 imagined for cesium
atoms in free fall and at absolute zero temperature \cite{aop14}.  This
specification is interpreted as defining a scale for a proper clock as
conceived by Einstein \cite{einstein05}.  Because no two clocks tick quite
alike, the reference for a live clock cannot be any realization of a live
clock, but must be a blackboard concept tied to a realization only to within
some tolerance.

\section{Logical synchronization}
\label{sec:3}
A live clock operates in a cycle of receiving unpredictable information from an
environment, storing that information in memory, computing a response, and
issuing that response to the environment.  The cycle has subcycles, and at the
finest level is composed of moments and moves of the clock-driven Turing
machine that makes up the live clock.  For a live clock to take part in
communication, its moments and moves have to be regulated to avoid the logical
conflict of a collision between writing into memory and reading from
memory. (In human terms this is the collision between trying to speak and
listen at the same time.)  To avoid this conflict,  the modified Turing machine is
driven by the adjustable clock through a cycle with two phases of moves and two
phases of moments, with reading from memory taking place in a phase separated
from a phase of writing into memory.

A cycle of the live clock corresponds to a unit interval of the readings of its
adjustable clock.  A reading of a live clock can be expressed in the form
$m.\phi_m$ where an integer $m$ indicates the count of cycles and $\phi_m$ is
the phase within the cycle.  We choose the convention that $-1/2<\phi_m\le
1/2$. (It is not necessary to think of the signals as points in time; it
suffices to think of a point reference within the signal.)

\subsection{Channels from one live clock to another}
To express the transmission of numbers from one live clock to another, we
follow Shannon in speaking of a communications channel; however we augment his
information-theoretic concept of a channel \cite{shannon48} with the live-clock
readings at the transmission and reception of character-bearing signals
\cite{aop14}.  Each character transmitted from a live clock $A$ to a live
clock $B$ is associated with a reading of live clock $A$ of the form $m.\phi_m$
at the transmission and with a reading of live clock $B$ of the form $n.\phi_n$
at the reception.  A channel from $A$ to $B$ includes a set of such pairs of
readings of the transmitting and the receiving live clocks.  The necessity of
avoiding a conflict between reading and writing imposes a constraint on the
phases of reception.

Restricting our attention to timing, we indicate a \emph{channel} from live
clock $A$ to live clock $B$, denoted $\abr$\,, as a set of pairs, each pair of
the form $(m.\phi_m,n.\phi_n)$.  The first member $m.\phi_m$ is an $A$-reading
at which live clock $A$ can transmit a signal and the second member $n.\phi_n$
is a $B$-reading at which live clock $B$ can register the reception of the
signal.  For theoretical purposes, it is convenient to define an
\emph{endlessly repeating channel} of the form
\begin{equation}
  \abr\, =\{(m+\ell j.\phi_{A,\ell},n+\ell k.\phi_{B,\ell})\},
\end{equation}
where $m,\,n,\,j,\,\text{and }k$ are fixed integers and $\ell$ ranges over all
integers. Again for theoretical purposes, we sometimes consider channels
for which the phases are all zero, in which case we may omit writing the
phases.

\begin{quote}
  \textbf{Proposition:} A character can propagate from one live clock to
  another only if the character arrives within the writing phase of the
  receiving live clock.
\end{quote}

When this phase constraint is met for a channel between a transmitting live
clock and a receiving live clock, we say the receiving live clock is
\emph{logically synchronized} to the transmitting live clock.  Logical
synchronization is analogous to the coordination between neighboring people in
a bucket brigade, or that between players tossing a ball back and forth, where
the arrival of the ball must be within a player's `phase of catching'.  In this
way the notion of a channel is expanded to include the clock readings that
indicate phases of signal arrivals that have to be controlled in order for the
logical synchronization of the channel to be maintained.  (While in many cases
the integers in clock readings that count cycles can be definitely specified,
the phases are never exactly predictable.)  We model the phase of writing at
which a live clock can receive a character as corresponding to
\begin{equation}\label{eq:ls}
  |\phi| < (1-\eta)/2,
\end{equation}
where $\eta$ (with 0 $< \eta < 1$) is a phase interval that makes room for
reading.

Logically synchronizing a channel means bringing about the condition
(\ref{eq:ls}) on phases at which signals arrive.  Once logical synchronization
is acquired for a set of channels, maintaining it typically requires more or less continually
adjusting the rates of ticking and the acceleration of the live
clocks, in order to steer the phases of arriving characters toward a suitable
reference pattern.  In the simplest case, this reference pattern entails zero
phases of reception.

Relations among readings of live clocks that contribute to a reference pattern
for a network include what we call echo counts, closely related to distances
defined by radar:
  \begin{quote}
  \textbf{Definition of echo count:} Suppose that at its reading $m.0$
  a live clock $A$ transmits a signal at to a live clock $B$,
  and the first signal that $B$ can transmit back to $A$ after
  receiving $A$'s signal reaches $A$ at $m'.\phi'$; then the quantity
  $m'.\phi'- m.0$ will be called the echo count $\Delta_{ABA}$ at $m$.
  \end{quote}

Although the concept of a channel is applicable without reference
to a spacetime manifold, in this paper we explore several questions of  possible
reference patterns of logically synchronized channels under the hypothesis of a
flat spacetime, so that we can speak of the period of a live clock as if it
could be determined by a proper clock of special relativity.  In
general that period can vary from tick to tick of the live clock \cite{aop14};
here however, we limit ourselves to the assumption of constant proper periods.
(Patterns of channels between live clocks in a curved spacetime are discussed
elsewhere \cite{aop14,04345}.)

\section{Logical synchronization where Einstein synchronization is precluded}
\label{sec:4}
Although channels linking live clocks are defined without any assumption of a
spacetime manifold, it is interesting to compare logical synchronization of
live clocks with the quite different synchronization defined by Einstein in
special relativity.  Einstein synchronization is stated in terms of
coordinates, and so to compare and contrast the logical synchronization of
channels linking live clocks with Einstein's synchronization, we assume for
this section a coordinate system that assigns flat spacetime coordinates to
ticks of live clocks (represented on the blackboard).  In special
relativity, the Sagnac effect precludes the Einstein synchronization of
neighboring live clocks attached to a rotating platform.  We review this
situation and contrast it with several cases in which channels between
neighboring live clocks can be logically synchronized, including theoretical
cases in which logical synchronization with zero phases at reception is
possible.

Consider $n$ live clocks $A_1,\ldots,A_n$ fixed to the nodes of a regular
polygon of $n$ sides, with $n \ge 3$ rotating in its plane about its center at
constant angular rate $\omega$, in a flat spacetime, with the center at rest in
some Lorentz frame, relative to which we use time and space coordinates.  Let
the radius of the polygon be $r$.  Let $T_{+}$ be the coordinate time duration
from transmission to reception by a nearest neighbor in the direction of rotation.
Let $T_-$ be the coordinate time duration from transmission to reception by
nearest neighbor in direction counter to rotation.  By symmetry, assume that
all the live clocks tick with a common period $p$ relative to the Lorentz
frame.

\subsection{Forward and backward propagation times}
Accounting for the rotation of a regular polygon of $n$ sides, one finds
relations among $T_+$, $T_-$, the radius $r$ of the polygon, angular rotation
rate $\omega$, and speed of light $c$:
\begin{eqnarray}
  cT_+/r &=& 2\sin\left(\frac{\pi}{n}+\frac{\omega T_+}{2}\right)\\
  cT_-/r &=& 2\sin\left(\frac{\pi}{n}-\frac{\omega T_-}{2}\right),
\end{eqnarray}
which implies a relation between $T_+$ and $T_-$ obtained by solving
these equations for $\omega$:
\begin{equation}\label{eq:ratio}
\frac{2}{T_+}\left[\sin^{-1}\left(\frac{cT_+}{2r}\right)-\frac{\pi}{n}\right]=
\frac{2}{T_-}\left[\frac{\pi}{n}-\sin^{-1}\left(\frac{cT_-}{2r}\right)\right] =
\omega.
\end{equation}
\begin{figure}[!h]
\includegraphics{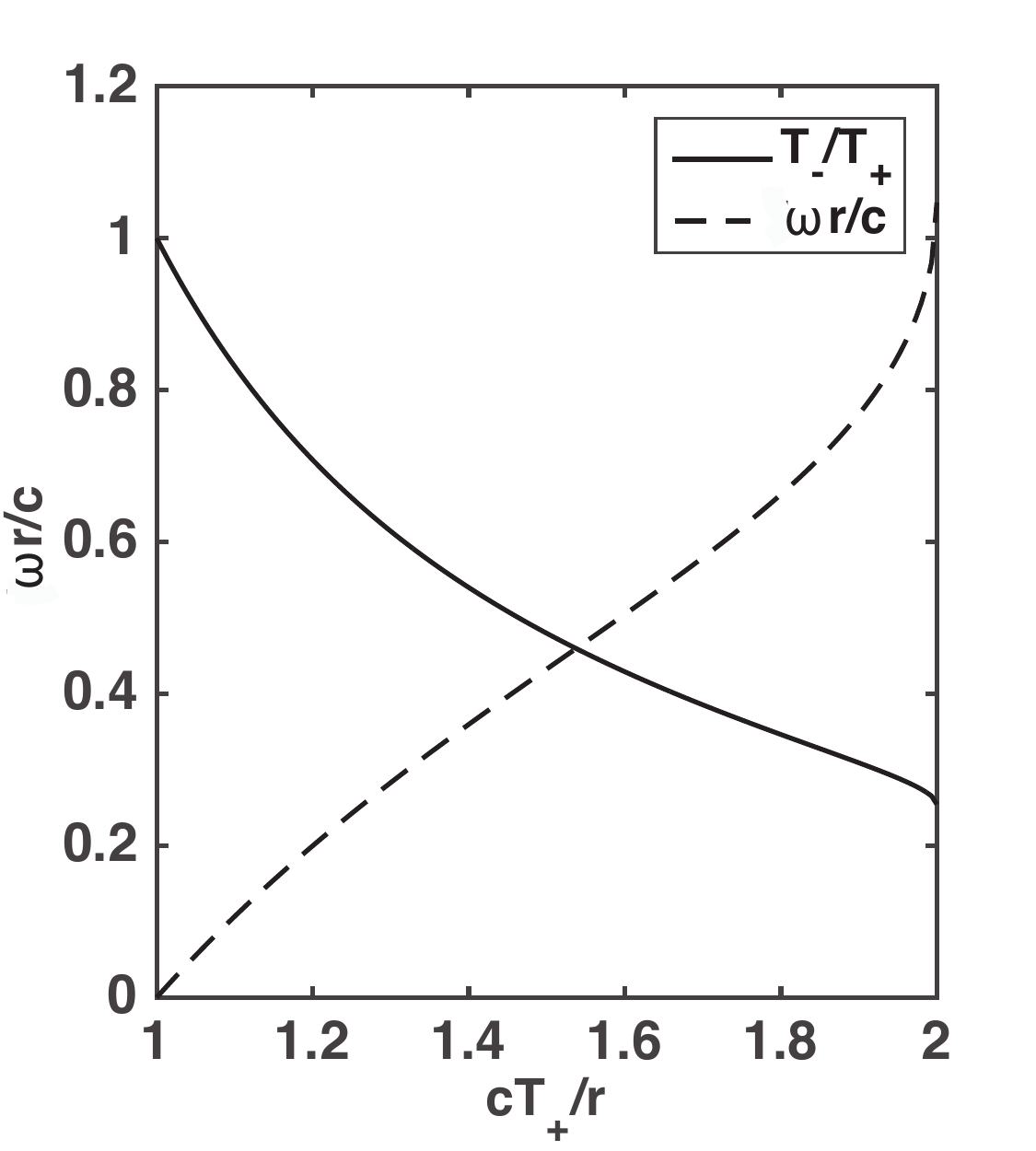}
\caption{\large{Ratio $T_-/T_+$ vs. $T_+$ and $\omega r/c$ vs. $T_+$ (dashed),
    from Eq.\ (\ref{eq:ratio}).}
\label{fig:1}}
\end{figure} 

  For the case of a rotating hexagonal arrangement of live clocks (so $n=6$),
  the ratio $T_-/T_+$ is plotted against $cT_+/r$ in Fig.\ \ref{fig:1}.  For a
  hexagonal arrangement in the absence of rotation, we have that $cT_+/r =1$,
  while $cT_+/r$ reaches a maximum value of 2 for the (hypothetical,
  superluminal) rotation of the arrangement that makes the light signal go
  diagonally across the hexagonal pattern.

\subsection{Einstein synchronization impossible}
Unless the angular velocity $\omega$ is zero,  $T_+>T_-$. For that reason if
rotating live clocks tick in coincidence relative to the Lorentz frame, they fail to
satisfy the Einstein criterion (\ref{eq:es}).  By shifting the ticks of each
clock in time, however, one can arrange for $A_2$ to be Einstein synchronous to
$A_1$, and so on through $A_n$ to $A_{n-1}$, but it is impossible to close the
loop to make $A_1$ synchronous to $A_n$.  Thus the Einstein synchronization
relation, which is transitive for clocks fixed to a Lorentz frame, is
\emph{not} a transitive relation for clocks on a rotating platform.  For a
rotating platform Einstein synchronization even in one rotational direction is
impossible.

\subsection{Logical synchronization for live clocks on a rotating platform}
For logical synchronization the situation is different, in one way more
restrictive, but in others less restrictive.  In particular, logical
synchronization allows for asymmetry in propagation times, that is, in cases
that $T_+ \ne T_-$.  For the cases considered in the following, assume all $n$
live clocks arranged at vertices of the rotating polygon have tick zero at
frame time $t=0$, and all continue to tick with the common period $p$.

\subsubsection{Case of low bandwidth based on tolerance of phasing}
By making the live-clock period $p$ sufficiently long, logical synchronization at
low bandwidth can operate merely by making the allowed phase interval for
reception longer than the duration for a signal to propagate from a live clock
to its nearest neighbor.  I.e. one arranges for the period $p$ of the live
clocks to be longer than the coordinate-time interval $T_+$, so that $T_+/p$ is
within the phase $(1-\eta)/2$ allowed for logical synchronization, per
Eq.\ (\ref{eq:ls}).  This of course limits bandwidth, which is proportional to
\begin{equation}
  1/p < \frac{1-\eta}{2 T_+}.
\end{equation}

\subsubsection{Case of one-way ring of logical synchronization with zero phases of reception}
More interesting are the cases of logical synchronization with zero phases.
Suppose $T_+=N_+p$ for $N_+$ a positive integer.  Then theoretically there can
be $n$ channels linking nearest neighbors in the direction of rotation:
\begin{equation}\label{eq:forward}
  \stackrel{\xrightarrow{\hspace*{1cm}}}{A_jA_{j+1}}=\{(k,k+N_+)|k \in \mathbb{Z}\},
\end{equation}
where $j  \in \{1,2,\ldots, n\}$ and we view $A_{n+1}$ as another name for $A_1$.
Given $T_+$ and any positive integer $N_+$, the necessary and sufficient
condition is met by a clock period $p =T_+/N_+$.

\subsection{Case of two-way ring of logical synchronization with zero phases of
  reception}

For two way logical synchronization of nearest-neighbor channels, there is a
restrictive relation between the radius of the polygon and the  angular
velocity of the platform.  Logically synchronized channels with zero receptive
phases from one live clock to its neighbor in the direction counter to rotation
have the form analogous to that of Eq. (\ref{eq:forward}):
\begin{equation}
  \stackrel{\xrightarrow{\hspace*{1cm}}}{A_jA_{j-1}}=\{(k,k+N_-)|k \in \mathbb{Z}\},
\end{equation}
where $j  \in \{1,2,\ldots, n\}$ and we view $A_{0}$ as another name for $A_n$.
Given $T_-$ and any positive integer $N_-$, the necessary and sufficient
condition is met by a clock period $p =T_-/N_-$.
Thus for two-way, 0-phase channels between nearest neighbors the period of the
live clocks has the necessary and sufficient condition
\begin{equation}
  p=T_+/N_+=T_-/N_-
\end{equation}
which requires that $T_-/T_+$, plotted in Fig.\ \ref{fig:1} for the case $n=6$,
be a rational number.

\subsection{Live clocks as tools of exploration}
In these examples, we assumed that a given angular velocity with respect to a
coordinate system is ``given'' and requires no action on their part.  More
interesting is the case in which the angular velocity is given to the live
clocks as a reference pattern, and the mission of the live clocks is to adjust
their accelerations and tick rates maneuver to obtain channels that correspond
to this reference angular velocity.  Another case is the exploration of
possible channels in order to measure a rotation rate, as in a Sagnac
interferometer.

\section{Discussion}\label{sec:5}
As a gentle illustration of logical synchronization, in the previous section we
invoked the familiar assumption of a coordinate system on a spacetime manifold
relative to which to describe an example of polygonal ring of rotating live
clocks with nearest neighbors linked by logically synchronized channels.
However, as already emphasized, the concept of logical synchronized channels
does not in itself make any assumption of a spacetime manifold, let alone a
coordinate system on that manifold.  Indeed the bringing about of a pattern of
logically synchronized channels provides a background against which to locate
events, tailored to one or another particular situation, and this background
provides some or all of the services asked of a coordinate frame.  That is, in
a network of logically synchronized channels linking live clocks, events are
located by their proximity to live-clock readings that, via these channels are
related to readings of the other live clocks of the network.  As discussed in
\cite{aop14}, gravitation affects the reference patterns of channels toward
which a network of live clocks can successfully steer, so that at high
precision the reference patterns are themselves hypotheses arrived at in part
by unpredictable guesswork, subject to revision in response to failures to come
within tolerable deviations from them.

The events that are critical to location are acts of transmitting and receiving
numbers, or, to put it a little more generally acts of transmitting and
receiving logical distinctions, such as the distinction between 0 and 1 or the
distinction between `yes' and `no'.  It is the communication of logical
distinctions in the face of an unpredictable environment that gives digital
computers their power, and indeed gives life the capacity to propagate through
the mechanisms of DNA replication.  The communication of logical distinctions
depends on regenerative amplification that reshapes signals to maintain logical
distinctions while allowing for tolerances in system components \cite{1639,aop05}.
Besides regenerative amplification, the communication of logical distinctions
requires phase management in the face of unpredictable environmental behavior.
The live clocks of a network function primarily not to tell ``time'' in the
sense of a spacetime coordinate, but to regulate the phasing needed for logical
synchronization.

We have stressed unpredictability of guesses that enter reference patterns, but
as discussed in \cite{aop14}, unpredictable events are also physical, as in the
detections by a photodetector.  While in some experiments, one accumulates such
detections passively to get an average rate of detections that can be related
to a probability, in other experiments, notably the operation of an atomic
clock, detections that cannot be individually predicted have to be responded to
promptly, and so enter the operation of clocks that realize the SI units of the
Hertz and the second.

There is lots left to do:
\begin{enumerate}
\item
  Can one retrieve some notion of a time coordinate that is available without
  the assumption of a spacetime manifold, based on tracing the implications of
  the relations between the transmissions of numbers and their receptions, as
  expressed by the channels of a network?
\item Regenerative amplification is found in biology, for example in the propagation of
  electrical spikes in nerve fibers \cite{nerve}.  Is phase management present
  in biology?
\item There are questions to ask and to answer concerning possibilities for
  patterns of channels among live clocks, whether in engineered systems or as
  found in living organisms.  To get a glimpse of the issue, for any live clock
  $A$, in principle there is a (likely variable) tick rate that will make it
  logically synchronized to signals from an arbitrary second live clock $B$,
  but the issue is not so simple if live clock $A$ wants to steer toward logical
  synchronization with a third live clock $C$ in addition to $B$.
\end{enumerate}

And there are bigger questions.  Quantum mechanics depends on coordinate
systems which, as we have seen, depend on clocks; however, the concept of a
coordinate system abstracts the clocks out of sight.  When we look into the
clocks and their communication by the transmission of number-carrying signals,
we find a situation readily described, as above, in terms of computer
engineering or its abstract Turing machines, without the use of quantum
language.  Realized clock networks depend on regenerative
amplification---thermodynamically non-reversible (even if logically reversible
\cite{bennett}) and outside of any graceful description in the language of
quantum theory.  Question: can some novel quantum-theoretic description
(involving decoherence?) represent networks of live clocks, or is quantum
mechanics irreducibly dependent on systems for locating events that are outside
its descriptive reach?

Finally there is the question of accepting or rejecting unpredictability as a
fundamental feature of life.  One often thinks of a coordinate system working
like rigid fences that organize a landscape, but if unpredictability is
pervasive, if the earth on which we stand shifts, which at present levels of
clock stability it always does, how is one to locate objects of interest? There
can be no rigid body on which to stand.  The application of physical laws that
underpin predictions requires number-carrying channels.  Channels operate in
the face of unpredictability that no law can shut out. When channels of a
network fail, as on occasion they do, the applications that depend on them
fragment.  Sometimes that fragmentation of a network calls us to search for a
different background pattern toward which to steer.  If needs to adjust our
reference patterns are in the cards, it is perhaps better to be nimble.
Recognizing unpredictability can be a first step toward that nimbleness.

\begin{acknowledgements}
We dedicate this paper to Howard Brandt, with whom, over decades, we have had
the benefit of wide-ranging discussions touching on predictability.
\end{acknowledgements}


\begin{thebibliography}{99}
\bibitem{tyler07} J.~M. Myers and F.~H. Madjid, ``Ambiguity in
  quantum-theoretical descriptions of experiments,'' in
  K. Mahdavi and D. Koslover, eds., \textit{Advances in Quantum
    Computation}, Contemporary Mathematics Series, vol.~482
  (American Mathematical Society, Providence, I, 2009),
  pp.\ 107--123; arXiv:1409.5678.
\bibitem{aop14} J. M. Myers and F. H. Madjid, ``Distinguishing between evidence
  and its explanations in the steering of atomic clocks,'' Annals of Physics
  350, 29--49 (2014); arXiv:1407.8020.
\bibitem{soffel03} M. Soffel et al., ``The IAU resolutions
  for astrometry, celestial mechanics, and metrology in the
  relativistic framework: explanatory supplement,'' The
  Astronomical Journal, \textbf{126}, 2687--2706 (2003).
\bibitem{einstein05} A. Einstein, ``Zur Elektrodynamik bewegter
  K\"orper,'' Annalen der Physik,  \textbf{17}, 891--921 (1905).
\bibitem{turing} A.~M. Turing, ``On computable numbers with an application to
  the Entscheidungsproblem,'' Proc.\ London Math.\ Soc., Series 2, \textbf{42},
  230--265 (1936).
\bibitem{04345} F. H. Madjid and J. M. Myers, ``Clocking in the face of
  unpredictability beyond quantum uncertainty,'' arXiv:1504.04345 (2015).
\bibitem{shannon48}C. E. Shannon, ``A mathematical theory of
  communication,'' The Bell System Technical Journal, Vol. 27,
  pp. 379--423, 623--656, July, October, 1948
\bibitem{aop05} F.~H. Madjid and J.~M. Myers, 
``Matched detectors as definers of force,'' 
\bibitem{1639} J. M. Myers and F. H. Madjid, ``Rhythms of Memory and
  Bits on Edge: Symbol Recognition as a Physical Phenomenon,''
  arXiv/1106.1639, 2011.
Ann.\ Physics \textbf{319} (2005), 251--273; arXiv:quant-ph/0404113..
\bibitem{nerve}J.~M. Myers, ``Modeling the effect of an external electric field on the
velocity of spike propagation in a nerve fiber,'' Phys.\ Rev.\ E, \textbf{60},
5918--5925 (1999).
\bibitem{bennett} C. H. Bennett, ``Logical reversibility of computation,'' IBM
  Journal of Research and Development, \textbf{17}, no. 6, 525--532 (1973).
\end{thebibliography}


\end{document}